\documentclass[aps,prl,reprint,superscriptaddress,
amsfonts,amssymb,amsmath,a4paper,floats,floatfix]{revtex4-1}
\usepackage{graphicx}
\usepackage{siunitx}
\usepackage[dvipsnames]{xcolor}

\begin{document}


\newcommand{\TDEG}{two-dimensional electron gas}


\title{Interactions and magnetotransport through spin-valley coupled Landau levels in monolayer MoS$_{2}$}

\author{Riccardo Pisoni}
	\affiliation{Solid State Physics Laboratory, ETH Z\"urich, 8093 Z\"urich, Switzerland}

\author{Andor Korm\'{a}nyos}
	\affiliation{Department of Physics, University of Konstanz, D-78464 Konstanz, Germany}
	\affiliation{Department of Physics of Complex Systems, E\"{o}tv\"{o}s Lor\'and University,
  P\'azm\'any P\'eter s\'et\'any 1/A, 1117 Budapest, Hungary}

\author{Matthew Brooks}
	\affiliation{Department of Physics, University of Konstanz, D-78464 Konstanz, Germany}
			
\author{Zijin Lei}
	\affiliation{Solid State Physics Laboratory, ETH Z\"urich, 8093 Z\"urich, Switzerland}

\author{Patrick Back}
	\affiliation{Institute of Quantum Electronics, Department of Physics, ETH Z{\"u}rich, 8093 Z{\"u}rich, Switzerland}

\author{Marius Eich}
	\affiliation{Solid State Physics Laboratory, ETH Z\"urich, 8093 Z\"urich, Switzerland}

\author{Hiske Overweg}
	\affiliation{Solid State Physics Laboratory, ETH Z\"urich, 8093 Z\"urich, Switzerland}

\author{Yongjin Lee}
	\affiliation{Solid State Physics Laboratory, ETH Z\"urich, 8093 Z\"urich, Switzerland}

\author{Peter Rickhaus}
	\affiliation{Solid State Physics Laboratory, ETH Z\"urich, 8093 Z\"urich, Switzerland}

\author{Kenji Watanabe}
	\affiliation{National Institute for Material Science, 1-1 Namiki, Tsukuba 305-0044, Japan}

\author{Takashi Taniguchi}
	\affiliation{National Institute for Material Science, 1-1 Namiki, Tsukuba 305-0044, Japan}

\author{Atac Imamoglu}
	\affiliation{Institute of Quantum Electronics, Department of Physics, ETH Z{\"u}rich, 8093 Z{\"u}rich, Switzerland}

\author{Guido Burkard}
	\affiliation{Department of Physics, University of Konstanz, D-78464 Konstanz, Germany}
	
\author{Thomas\ Ihn}
	\affiliation{Solid State Physics Laboratory, ETH Z\"urich, 8093 Z\"urich, Switzerland}
	
\author{Klaus\ Ensslin}
	\affiliation{Solid State Physics Laboratory, ETH Z\"urich, 8093 Z\"urich, Switzerland}

\date{\today}


\begin{abstract}
The strong spin-orbit coupling and the broken inversion symmetry in monolayer transition metal dichalcogenides (TMDs) results in spin-valley coupled band structures. Such a band structure leads to novel applications in the fields of electronics and optoelectronics. Density functional theory calculations as well as optical experiments have focused on spin-valley coupling in the valence band. Here we present magnetotransport experiments on high-quality n-type monolayer molybdenum disulphide (MoS$_{2}$) samples, displaying highly resolved Shubnikov-de Haas oscillations at magnetic fields as low as \SI{2}{T}. We find the effective mass \SI{0.7}{m_{e}}, about twice as large as theoretically predicted and almost independent of magnetic field and carrier density. We further detect the occupation of the second spin-orbit split band at an energy of about \SI{15}{meV}, i.e. about a factor $5$ larger than predicted. In addition, we demonstrate an intricate Landau level spectrum arising from a complex interplay between a density-dependent Zeeman splitting and spin and valley-split Landau levels. These observations, enabled by the high electronic quality of our samples, testify to the importance of interaction effects in the conduction band of monolayer MoS$_{2}$.
\end{abstract}

\maketitle

Monolayer transition metal dichalcogenides (TMDs) such as MoS$_{2}$, MoSe$_{2}$, WS$_{2}$ and WSe$_{2}$ are two-dimensional (2D) semiconductors with band extrema at the corners (K, K$'$-points) of the first Brillouin zone~\cite{xiao_coupled_2012}. Due to the strong spin-orbit coupling the spin degeneracy in the K and K$'$ valleys is lifted, with opposite spin polarization normal to the layer plane in opposite valleys (see Fig.~\ref{fig:fig2}, inset). This peculiar band structure with coupled spin and valley degrees of freedom results in an anomalous Landau level (LL) structure~\cite{li_unconventional_2013,wang_valley-_2017,kormanyos_landau_2015}.
Theoretical proposals predict the formation of LLs under the influence of a perpendicular magnetic field that are arranged differently from those in conventional semiconductor quantum wells and graphene~\cite{wang_valley-_2017}. Magnetotransport measurements have recently been performed in monolayer WSe$_{2}$, MoSe$_{2}$ and bilayer MoS$_{2}$ revealing two-fold degenerate LLs, large effective masses and carrier density dependent Zeeman splitting~\cite{fallahazad_shubnikovchar21haas_2016,movva_density-dependent_2017,gustafsson_ambipolar_2018,larentis_large_2018,lin_probing_2018}. Previous works on thicker MoS$_{2}$, MoSe$_{2}$ and WSe$_{2}$ devices have measured the electron LLs structure at the Q and Q$'$ conduction band minima, 
showing the thickness dependence of the band structure in 2D TMDs~\cite{wu_evenodd_2016,pisoni_gate-defined_2017}. Here we focus on single layer MoS$_{2}$ where for low electron densities electrons clearly reside at the K-K$'$ minima of the bandstructure.

Here we report transport measurements in high mobility dual-gated monolayer MoS$_{2}$ under a perpendicular magnetic field. Our devices show ohmic contacts at temperatures as low as $T\approx\SI{100}{mK}$ allowing us to uncover signatures of so far not reported rich interplay of strong spin-orbit coupling and electron-electron interactions. Shubnikov-de Haas (SdH) oscillations appear already at magnetic fields $B\approx\SI{2}{T}$ at a temperature of $T\approx\SI{100}{mK}$. From the temperature dependence of the SdH oscillations we measure an electron effective mass of $\approx\SI{0.7}{m_{e}}$, compared to a value of $\SI{0.4}{m_{e}}$ predicted by Density Functional Theory (DFT) calculations~\cite{kormanyos_spin-orbit_2014,kormanyos_monolayer_2013,wang_electronics_2012,liu_three-band_2013}. By increasing the electron density, we observe an interplay between even and odd filling factor sequences explained qualitatively by a density-dependent effective $g$-factor, similar to observations in p-doped WSe$_2$ and n-doped MoSe$_2$ monolayers~\cite{movva_density-dependent_2017,larentis_large_2018}. At electron densities $>\SI{4}\times{10^{12}}{~\mathrm{cm^{-2}}}$, corresponding to a Fermi energy $>\SI{15}{meV}$, the upper spin-orbit split bands start to be populated and the complex LL structure of the different valley-spin polarized bands is observed. We give evidence of intricate physics beyond the single particle picture that was employed to explain the experimental results in previous works~\cite{fallahazad_shubnikovchar21haas_2016,movva_density-dependent_2017,gustafsson_ambipolar_2018,larentis_large_2018,lin_probing_2018}.

\begin{figure}
\includegraphics[width=1\columnwidth]{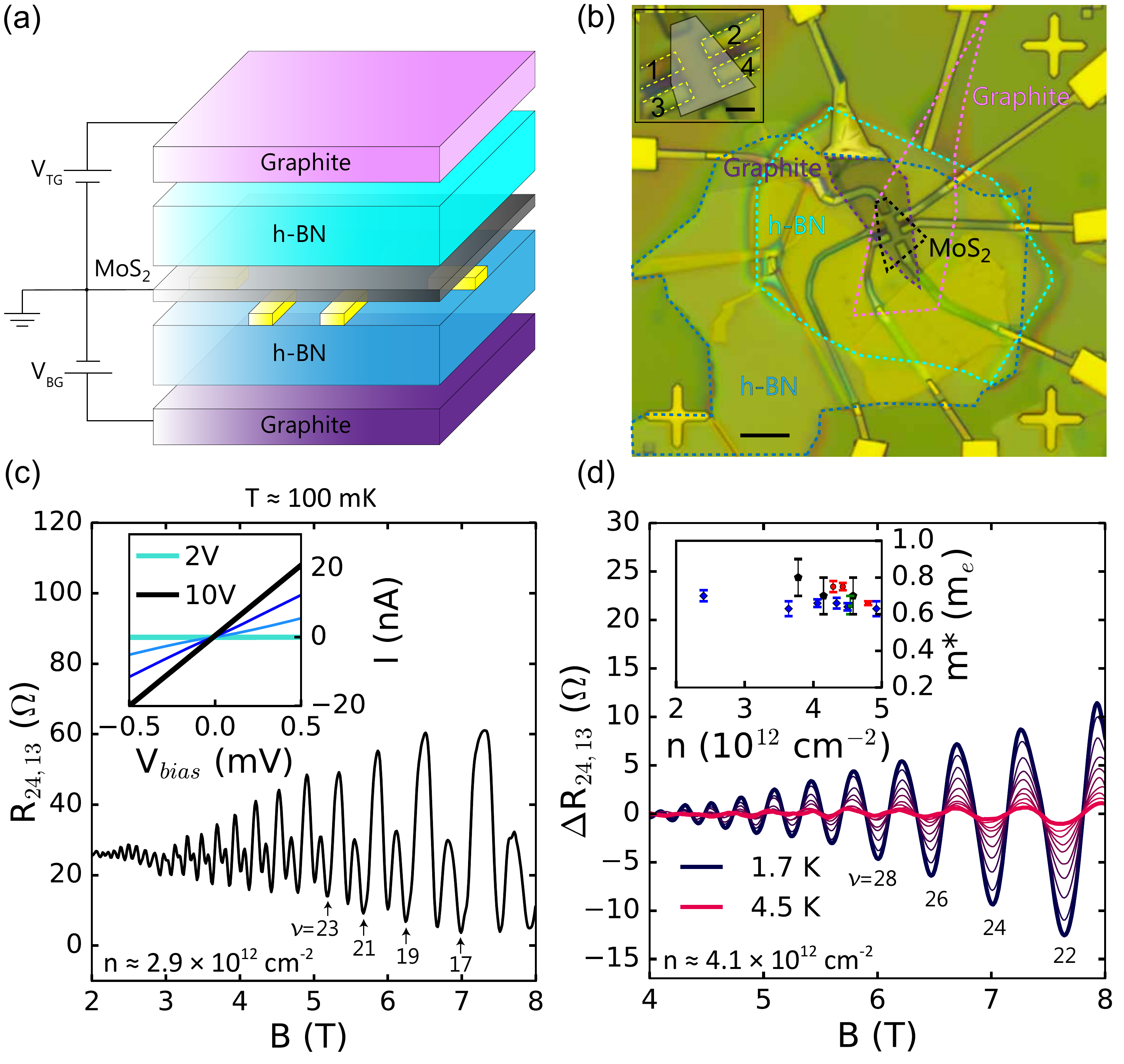}
\caption{\label{fig:fig1}
(a)~Device schematic. A single layer MoS$_{2}$ is encapsulated between two layers of hBN. Graphite flakes are used as bottom and top gates. Ti/Au electrodes are evaporated on top of the bottom hBN before the MoS$_{2}$ layer is transferred.
(b)~Optical micrograph of the sample. The MoS$_{2}$ flake is highlighted with black dashed lines (scale bar is  \SI{10}{\micro m}). Inset: Ti/Au contacts to the MoS$_{2}$ flake are numbered 1 - 4 (Scale bar is \SI{2}{\micro m}).
(c)~Four-terminal resistance $R_{24,13}$ as a function of $B$ at $V_\mathrm{BG} = \SI{-2.2}{V}$, $n_{\mathrm{SdH}}\approx\SI{2.9}\times{10^{12}}{~\mathrm{cm^{-2}}}$ and $T \approx\SI{100}{mK}$. SdH oscillations appear at $B \approx\SI{2}{T}$. We observe a predominantly odd filling factor sequence $\nu = 17, 19, 21, 23$. Inset: Linear $I-V_\mathrm{bias}$ traces as a function of $V_\mathrm{TG}$ at $T \approx\SI{100}{mK}$. The Ohmic contact regime is achieved for $V_\mathrm{TG}> \SI{2}{V}$. 
(d)~SdH oscillations as a function of the magnetic field for different temperatures at $V_\mathrm{BG} =\SI{-1.5}{V}$, $n_{\mathrm{SdH}}\approx\SI{4.1}\times{10^{12}}{~\mathrm{cm^{-2}}}$. An Even filling factor sequence $\nu = 22, 24, 26, 28$ is measured. Inset: effective mass $m^{\ast}$ calculated for the four samples as a function of electron density. Blue, red, green and black markers correspond to samples A, B, C and D, respectively.
}
\end{figure}
\begin{figure}
\includegraphics[width=1\columnwidth]{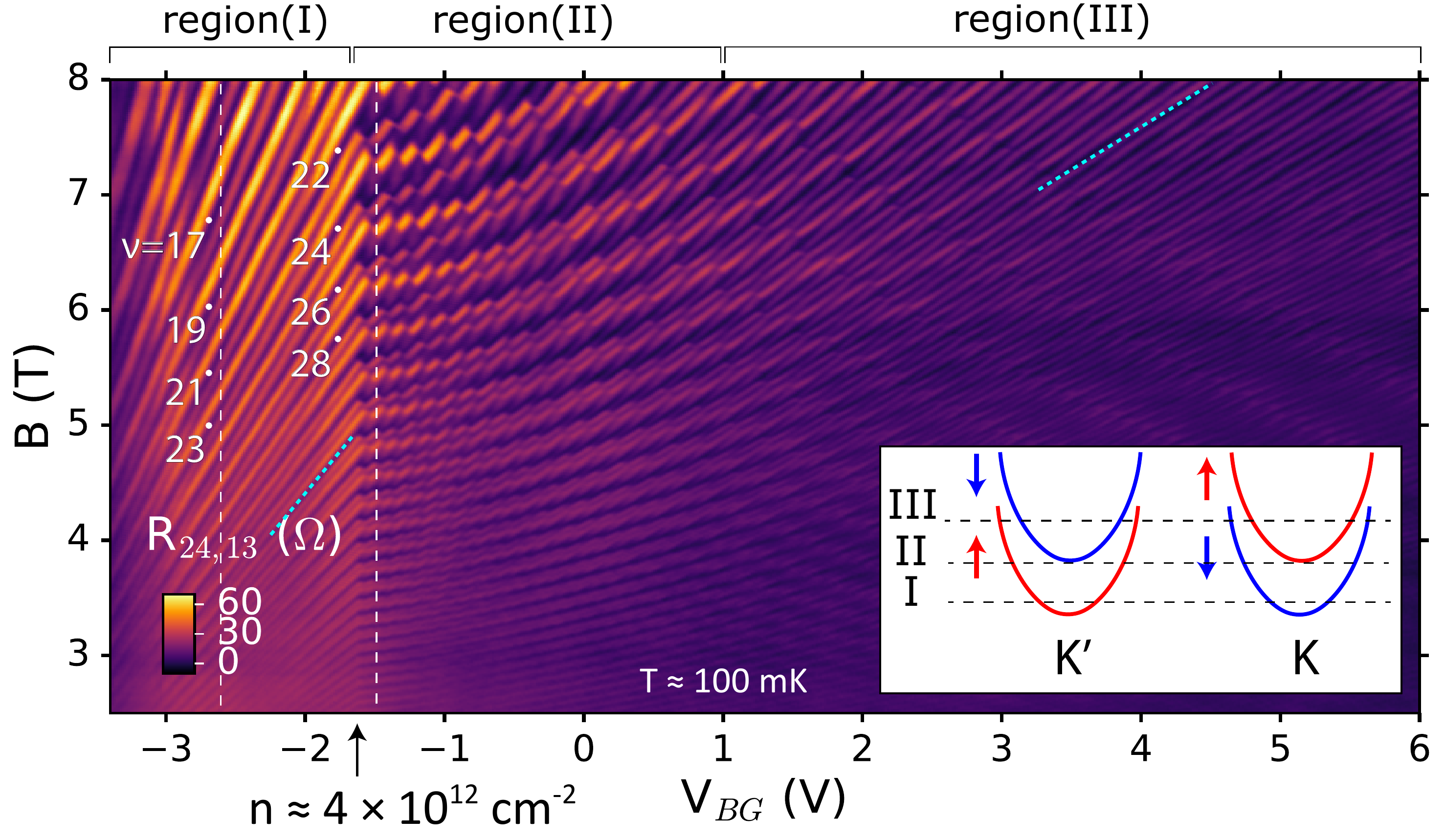}
\caption{\label{fig:fig2}
Four-terminal resistance $R_{24,13}$ as a function of $V_\mathrm{BG}$ and magnetic field at $T \approx \SI{100}{mK}$. We observe a pronounced change in the slope (cyan dashed lines) of the Landau fan diagram at $V_\mathrm{BG}\approx\SI{-1.6}{V}$, $n_{\mathrm{SdH}}\approx\SI{4}\times{10^{12}}{~\mathrm{cm^{-2}}}$ (black arrow). The Landau fan diagram can be divided in three different regions: (I)  $V_\mathrm{BG}< \SI{-1.6}{V}$; (II) $\SI{-1.6}{V} < V_\mathrm{BG}< \SI{1}{V}$; (III) $V_\mathrm{BG}> \SI{1}{V}$. Left and right white dashed lines correspond to the line cuts presented in Fig.~\ref{fig:fig1}(c) and Fig.~\ref{fig:fig1}(d), respectively. Inset: Sketch of the conduction band minima at the K and K$'$ points in the first Brillouin zone of monolayer MoS$_{2}$. Due to the strong spin-orbit interaction the spin degeneracy is lifted and spin and valley degrees of freedom are locked. Black dashed lines represent the Fermi energy corresponding to regions (I), (II) and (III), respectively.
}
\end{figure}

High mobility TMD field effect devices were fabricated using a van der Waals heterostructure platform~\cite{pisoni_gate-defined_2017,pisoni_gate-tunable_2018}. A schematic of the device is shown in Fig.~\ref{fig:fig1}(a). Monolayer MoS$_{2}$ flakes were encapsulated between two hexagonal boron nitride (hBN) layers and graphite flakes serve as top and bottom gates.  We fabricated and measured four monolayer MoS$_{2}$ samples, labelled A, B, C and D, which show consistent behaviour. We will mainly discuss sample A here. Data from samples B, C and D are presented in the Supplementary.
Fig.~\ref{fig:fig1}(b) shows the optical micrograph of sample A with the MoS$_{2}$ flake outlined in black. The inset of Fig.~\ref{fig:fig1}(b) displays a sketch of the MoS$_{2}$ flake where the ohmic contacts are numbered from 1 to 4. Contacts 2 and 4 are used for current injection and extraction, contacts 1 and 3 serve as voltage probes. The top and bottom hBN outlined in cyan and blue, respectively, serve as dielectric layers to insulate the conducting MoS$_{2}$ from the top and bottom graphite gates. We use graphite as a gate electrode because it provides an atomically flat surface and a uniform potential landscape for the MoS$_{2}$ layer~\cite{pisoni_gate-defined_2017,bretheau_tunnelling_2017,hiske,wang_electronic_2015}.

The high electron mobility and low contact resistances allow us to investigate quantum transport phenomena in single-layer MoS$_{2}$ using standard lock-in techniques at \SI{31.4}{Hz}. 
All measurements presented here are performed at $V_\mathrm{TG}=\SI{8}{V}$ in order to ensure Ohmic behaviour of the contacts at low temperatures. In Fig.~\ref{fig:fig1}(c) we present the four terminal resistance $R_{24,13}$ as a function of magnetic field $B$ at $V_\mathrm{BG}=\SI{-2.2}{V}$, $n_{\mathrm{SdH}}\approx\SI{2.9}\times{10^{12}}{~\mathrm{cm^{-2}}}$, and $T \approx\SI{100}{mK}$ (left vertical dashed line in Fig.~\ref{fig:fig2}). SdH oscillations start at $B\approx\SI{2}{T}$ yielding a lower bound for the quantum mobility of $\approx\mathrm{5,000}~\mathrm{cm^{2}/Vs}$. The electron density is determined from the SdH oscillations according to $n_{\mathrm{SdH}} = (e/h)(1/\Delta(1/B))$, where $\Delta (1/B)$ is the period of the SdH oscillations in $1/B$. At  $n_{\mathrm{SdH}}\approx\SI{2.9}\times{10^{12}}{~\mathrm{cm^{-2}}}$ we measure an alternating sequence of deeper and shallower minima corresponding to odd and even filling factors, respectively, meaning that the Landau levels of the K and K$'$ valleys are no longer degenerate. The inset of Fig.~\ref{fig:fig1}(c) displays the $I-V_\mathrm{bias}$ traces as a function of $V_\mathrm{TG}$ at $V_\mathrm{BG} = \SI{0}{V}$ and $T \approx \SI{100}{mK}$. The linearity of the $I-V_\mathrm{bias}$ curves for $V_\mathrm{TG}>\SI{2}{V}$ indicates the regime of a good ohmic contact at low temperatures.

In order to determine the effective mass we measure in Fig.~\ref{fig:fig1}(d) the four-terminal resistance $\Delta R_{24,13}$ with a smooth background subtracted as a function of $B$ at various elevated temperatures ranging from \SI{1.7} to \SI{4.5}{K} at $V_\mathrm{BG}=\SI{-1.5}{V}$, $n_{\mathrm{SdH}}\approx\SI{4.1}\times{10^{12}}{~\mathrm{cm^{-2}}}$ (right vertical dashed line in Fig.~\ref{fig:fig2}). We observe the sequence of even filling factors $\nu = 22,~24,~26,~28$ since the splitting of the K and K$'$ valleys is not resolved for these elevated temperatures. From the $T$-dependence of the SdH oscillation amplitudes we extract the electron effective mass $m^{\ast}$ by fitting  $\Delta R_{24,13}$ to $x/sinh(x)$, where $x = 2\pi^{2}k_{B}T/\hbar\omega_{c}$ and $\omega_{c} = eB/m^{\ast}$ is the cyclotron frequency (see Supplementary)~\cite{ando_electronic_1982,isihara_density_1986,pudalov_probing_2014}. In the inset of Fig.~\ref{fig:fig1}(d) we present $m^{\ast}$ at various electron densities $n_{\mathrm{SdH}}$ for the four different samples. For samples A and C we calculate the density-averaged mass ${m^{\ast}/m_{e}}=0.65 \pm 0.04$ where $m_{e}$ is the electron rest mass. For sample B, $m^{\ast}/m_{e} = 0.75 \pm 0.03$. Sample D shows $0.7 \leq m^{\ast}/m_{e}\leq0.8$ with larger error bars compared to the other three samples due to a less precise temperature calibration. No obvious dependence of the mass on $n_{\mathrm{SdH}}$ or $B$ is observed~\cite{zhang_density-dependent_2005,attaccalite_correlation_2002}. These $m^{\ast}$ values are larger than those of DFT studies which predict $m^{\ast}/m_{e}\approx0.4$  for single layer MoS$_{2}$~\cite{kormanyos_spin-orbit_2014,kormanyos_monolayer_2013,wang_electronics_2012,liu_three-band_2013}.

In Fig.~\ref{fig:fig2} we present an overview of the four-terminal magneto-resistance $R_{24,13}$ (color scale) over a wide range of $V_\mathrm{BG}$ and $B$ applied perpendicularly to the sample at $T\approx\SI{100}{mK}$. There are three qualitatively different regions which we discuss in the following. The first region (I) corresponds to $V_\mathrm{BG}<\SI{-1.6}{V}$, the second region (II) to $\SI{-1.6}{V}<V_\mathrm{BG}<\SI{1}{V}$ and the third region (III) to $V_\mathrm{BG}>\SI{1}{V}$. The black dashed lines in the inset of Fig.~\ref{fig:fig2} indicate the Fermi energies corresponding to regions (I), (II) and (III).

\begin{figure}
\includegraphics[width=1\columnwidth]{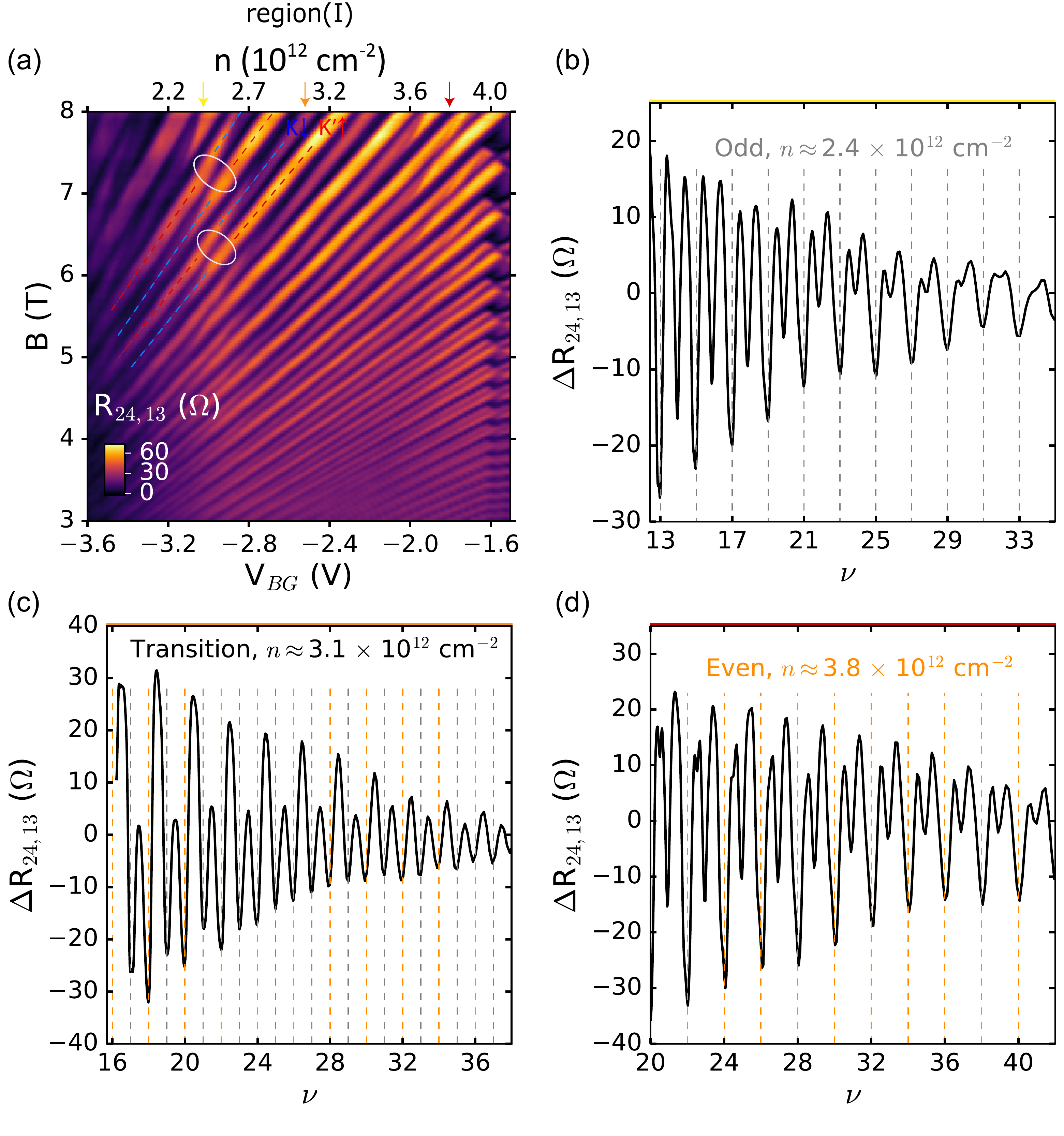}
\caption{\label{fig:fig3}
(a)~Region (I). $R_{24,13}$ as a function of electron density $n_{\mathrm{SdH}}$ and magnetic field at $T \approx \SI{100}{mK}$. Two sets of LLs corresponding to the K spin-down and K$'$ spin-up valleys can be distinguished (blue and red dashed lines, respectively). Anticrossings between not fully spin polarized LLs appear (white circles).
(b-d)~Four terminal resistance $\Delta R_{24,13}$, after subtracting a smooth background, as a function of LL filling factors at different electron densities. By increasing the electron density we observe an interplay between predominantly odd and predominantly even filling factors sequences (yellow, orange and red arrows in \textbf{a}, respectively).
}
\end{figure}

We now discuss region (I) of the Landau fan diagram shown in Fig.~\ref{fig:fig2}.
Fig.~\ref{fig:fig3}(a) shows a blow-up of the four-terminal resistance $R_{24,31}$ in this region as a function of $V_\mathrm{BG}$ and $B$ at $T\approx\SI{100}{mK}$. We observe two sets of LLs with different amplitudes that we attribute to the valley K spin-down and K$'$ spin-up LLs. 
In Fig.~\ref{fig:fig3}(b-d) we show the SdH oscillations for three representative electron densities. 
Both for $n_{SdH}=\SI{2.4}\times{10^{12}}{~\mathrm{cm^{-2}}}$ (Fig.~\ref{fig:fig3}(b)) and  $n_{SdH}=\SI{3.8}\times{10^{12}}{~\mathrm{cm^{-2}}}$ (Fig.~\ref{fig:fig3}(d)) one can observe an alternating 
sequence of deeper (primary) and shallower (secondary) minima. While for the lower density the primary minima are at odd filling
factors $\nu$, for the higher density they are at even $\nu$. For the transition density $n_{SdH}=\SI{3.1}\times{10^{12}}{~\mathrm{cm^{-2}}}$ (Fig.~\ref{fig:fig3}(c)) the minima
at even and odd filling factors are approximately equally deep. This means that 
by tuning the electron density we observe a transition from a predominantly odd to a predominantly 
even filling factor sequence.

For the considered electron densities electron-electron interactions are expected to play
a significant role~\cite{larentis_large_2018,movva_density-dependent_2017}, similar to other multi-valley two-dimensional systems~\cite{okamoto_spin_1999,shashkin_indication_2001,vakili_spin_2004}.

The interaction strength can be characterized by the dimensionless Wigner-Seitz radius 
$r_s=1/(\sqrt{\pi\,n_\mathrm{e}}a^{\ast}_\mathrm{B})$, where $a^{\ast}_\mathrm{B}=a_\mathrm{B}(\kappa m_\mathrm{e}/m^{\ast})$ is the 
effective Bohr radius, $\kappa$ the dielectric constant and $a_\mathrm{B}$ the Bohr radius. 
For the regime $\SI{-3.2}{V}<V_\mathrm{BG}<\SI{-1.6}{V}$ we estimate that $r_s=9.8 - 7.5$, placing the system in a regime where 
interactions are important. Qualitatively, the observations can be explained 
by an extended single particle picture, where electron-electron interaction effects are accounted for by assuming 
i) $n_{\mathrm{SdH}}$-\emph{dependent} valley $g$-factor $g_\mathrm{vl}^{}$, 
and ii) in good approximation $n_{\mathrm{SdH}}$-\emph{independent} effective mass $m^{\ast}$.  

For data taken at \SI{1.7}{K} there are regimes where only even or odd filling factors are visible (see Supplementary). A model suggested in the literature~\cite{movva_density-dependent_2017,larentis_large_2018} based on a density dependent $g$-factor is in good agreement with these data (see Supplementary). Such a model will necessarily lead to a situation where neighbouring Landau levels become accidentally degenerate and the corresponding SdH minimum will disappear. Our low temperature data shown in Fig.~\ref{fig:fig3}(b-d)  indicate however, that neighbouring Landau levels are never degenerate, demonstrating the limitations of such a simple model. Extreme cases of such anti-crossings at $T\approx\SI{100}{mK}$ are indicated with white circles in Fig.~\ref{fig:fig3}(a). These anticrossings, which happen for approximately integer value of the ratio of valley Zeeman energy with respect to cyclotron energy $E_{vz}/E_{c}$,  cannot be explained in a  single particle picture where the LLs 
in the K  and K$'$ valleys are assumed to have out-of-plane (i.e., parallel to the magnetic field) and 
orthogonal spin-polarization,  since this  would imply that they should  cross. Instead, these anticrossings can arise as a result of electron-electron interaction effects, that mix single-particle LLs of opposite spin and lead to not fully spin polarized LLs.

Continuing with region (II) in Fig.~\ref{fig:fig2}, at $V_\mathrm{BG} = \SI{-1.6}{V}$, $n_{\mathrm{SdH}}\approx\SI{4}\times{10^{12}}{~\mathrm{cm^{-2}}}$ (black arrow), we observe two important 
changes in the SdH oscillations. First, as shown in the Supplementary, there is another even to odd transition in the SdH oscillation minima like in region (I) with the same interpretation.  
Second, as indicated in Fig.~\ref{fig:fig2} with cyan dashed lines [one in region (I), the other in region (III)], there is a sudden change in the slope of the SdH minima related to constant filling factor by about a factor of two in the Landau fan diagram at the left edge of region (II). A factor of two is expected when the density of states doubles.
The Hall mobility $\mu$ and the four terminal resistance $R_{24,13}$ at zero magnetic field exhibit a pronounced 
change in slope at the same point (Supplementary). We attribute these observations to the occupation of the 
upper spin-orbit split K and K$'$ valleys in the conduction band of monolayer MoS$_2$ , as sketched in the inset of Fig.~\ref{fig:fig1}(c). 

\begin{figure}
\includegraphics[width=1\columnwidth]{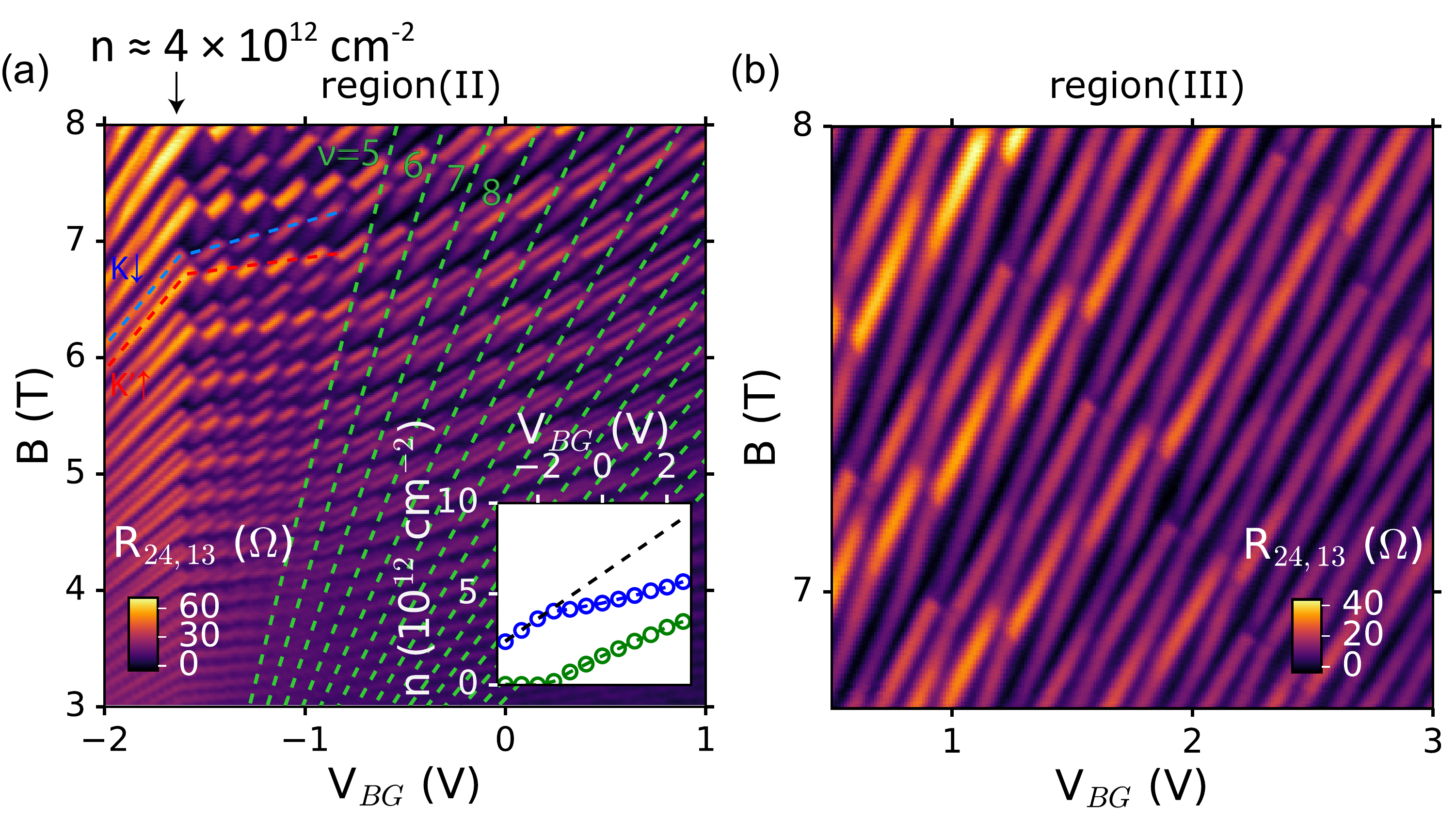}
\caption{\label{fig:fig4}
(a)~Region (II). $R_{24,13}$, as a function of $V_\mathrm{BG}$ and magnetic field at $T \approx\SI{100}{mK}$. Inset: electron density as a function of $V_\mathrm{BG}$. Black dashed line indicate the total electron density determined from the capacitor model. Blue and green markers represent the lower and upper spin-orbit split bands electron densities, respectively. Green dashed lines in Fig.~\ref{fig:fig4}(a) indicate the Landau fan diagram originating from the upper spin-orbit split bands.
(b)~Region (III). $R_{24,13}$ as a function of $V_\mathrm{BG}$ and magnetic field at $T \approx\SI{100}{mK}$. Multiple anticrossings through spin-valley coupled LLs are observed.
}
\end{figure}

A zoom into region (II) is shown in Fig.~\ref{fig:fig4}(a). As we show in the inset of Fig.~\ref{fig:fig4}(a), the difference between the total electron density (black dashed line) and that of the lower spin-orbit split bands (blue circles) increases linearly as a function of $V_\mathrm{BG}$ for $n_{\mathrm{SdH}}\approx\SI{4}\times{10^{12}}{~\mathrm{cm^{-2}}}$. The ``missing'' electron density (green circles) leads to a calculated additional Landau fan (green dashed lines in Fig.~\ref{fig:fig3}(a)) which is compatible with the appearance of the intermittent shifts of the SdH maxima in this region.

The threshold electron density where the slope change occurs is $n_{\mathrm{SdH}}\approx\SI{4}\times{10^{12}}{~\mathrm{cm^{-2}}}$. Assuming a 2D density of states 
$DOS = m^{*}/\pi \hbar^{2}$ implying a two-fold degeneracy and using the experimentally determined effective mass, we calculate the Fermi energy to be $E_F \approx \SI{15}{meV}$, 
which gives us an estimate of the intrinsic spin-orbit interaction $2 \Delta_{cb}^{*}$ for K-valley electrons in monolayer MoS$_2$. 
This value of  $2 \Delta_{cb}^{*}$ is about a factor of five larger than the results of DFT band 
structure calculations~\cite{kormanyos_spin-orbit_2014}. We note that a similar (albeit smaller) enhancement of $2 \Delta_{cb}^{*}$ with respect to theoretical calculations 
was also observed~\cite{larentis_large_2018} for monolayer MoSe$_2$. 
This apparent enhancement of the spin splitting of the bands might be due to an exchange interaction driven band renormalization. 
Although the upper spin-orbit split bands start to be filled, we found that, at $T\approx\SI{1.7}{K}$ the measurements can still 
be fitted nicely (see Supplementary) assuming that only the lower spin-orbit split bands give visible contributions to the SdH oscillations. The effect of the LLs corresponding 
to the upper spin-orbit split  bands become apparent at lower temperatures where the distinctive ``wavyness`` of the bright lines (Fig.~\ref{fig:fig4}(a)) suggests that LLs corresponding to the lower spin-orbit split bands
are affected by the LLs originating from the upper spin-orbit split bands.

We also note that, as discussed in the Supplementary, photoluminescence measurements exhibit, at $n_{\mathrm{SdH}}\approx\SI{4}\times{10^{12}}{~\mathrm{cm^{-2}}}$, a third peak, at lower energy than that of the exciton and the attractive polaron\cite{sidler_fermi_2017} peaks, indicating the emergence of a new emission channel. 


Fig.~\ref{fig:fig4}(b) shows a zoom-in of Fig.~\ref{fig:fig2} for region (III) at $T\approx\SI{100}{mK}$. 
We observe the appearance of anticrossings, a signature of split spin-valley coupled LLs originating from the lower and upper spin-orbit split bands. These observations indicate again that a standard single particle picture for the description of Landau levels is insufficient. It is remarkable that the observed anticrossings seem independent of which spin or/and valley states are involved. No obvious selection rules can be observed. Experimentally we find that the level anti-crossings at a magnetic field of \SI{6}{T} can be resolved below a temperature of about \SI{500}{mK}. This corresponds to an estimated interaction energy scale of  $4k_\mathrm{B}T\approx \SI{170}{\micro eV}$. In comparison the single-particle LL splitting $\hbar \omega_\mathrm{c}$ calculated with an effective mass of \SI{0.65}{m_{e}} at \SI{6}{T} is \SI{1}{meV}. We see that the interaction energy is a significant fraction of the Landau level spacing. The disorder-limited energy resolution for LL energy gaps in our experiment must be well below the estimated interaction energy. This in turn is in rough agreement with the quantum mobility of $\mathrm{5,000}~\mathrm{cm^{2}/Vs}$, which leads to an upper bound for the characteristic disorder energy of \SI{180}{\micro eV} if we assume the experimentally deduced effective mass of \SI{0.65}{m_{e}}.

Compared to standard semiconductor 2D electron gases such as those in high-mobility AlGaAs heterostructures, where most effects can be quantitatively explained within a single particle-model with the inclusion of exchange effects for small odd filling factors, the data presented here indicates that the formation of Landau levels in monolayer MoS$_{2}$ is governed if not dictated by the combination of both spin-orbit and carrier-carrier interactions.

Quantum Hall ferromagnetism~\cite{poortere_resistance_2000,jungwirth_magnetic_1998} is relevant for small filling factors ($< 5$) and it becomes less pronounced for larger filling factors since a possible overall spin/valley polarization decreases. For our experiments we deal with large filling factors ($> 20$) and the observed anti-crossings appear to be independent of filling factor. We conclude that exchange enhancement of the $g$-factor which causes quantum Hall ferromagnetism is not relevant here.


In conclusion, we fabricated high mobility dual-gated single layer MoS$_{2}$ devices using a van der Waals heterostructure platform with quantum mobilities as high as $\mu\approx\mathrm{5,000}~\mathrm{cm^{2}/Vs}$. The temperature dependence of the SdH oscillations reveals an electron effective mass of $\approx\SI{0.7}{m_{e}}$. We are able to measure and resolve the LL structure of the lower spin-orbit split K and K$'$ valleys. At $n_{\mathrm{SdH}}\approx\SI{4}\times{10^{12}}{~\mathrm{cm^{-2}}}$, we observe the occupation of the upper spin-orbit split K and K$'$ valleys, thus estimating $2\Delta_{cb}^{*}\approx\SI{15}{meV}$. At higher electron densities we observe the appearance of multiple sets of LLs originating from the upper and lower spin-orbit split K and K$'$ valleys. Interaction effects of valley and spin polarized LLs, at elevated temperatures consistent with a density-dependent $g$-factor, are observed in the experiments.
Measurements of the LL structure of monolayer MoS$_{2}$ has been hindered to-date by high contact resistances and low sample mobilities. Our results demonstrate the subtle and unconventional conduction band Landau level structure of monolayer MoS$_{2}$, where strong spin-orbit interaction meets strong electron-electron interactions. This indicates the presence of rich, novel and so far unpredicted physics possibly beyond that expected from single-particle considerations. These prospects bear relevance also for related TMD materials, such as MoSe$_{2}$, WS$_{2}$ and WSe$_{2}$.

\begin{acknowledgments}
We thank Emanuel Tutuc, Beat Br{\"a}m, Ovidiu Cotlet, Matija Karalic and Giorgio Nicol\'i for fruitful discussions. We thank Peter M{\"a}rki, Erwin Studer, as well as the FIRST staff for their technical support. We acknowledge financial support from ITN Spin-NANO Marie Sklodowska-Curie grant agreement no. 676108, the Graphene Flagship and the National Center of Competence in Research on Quantum Science and Technology (NCCR QSIT) funded by the Swiss National Science Foundation. 
A.K. was supported by the National Research Development and Innovation Office of Hungary within the Quantum Technology National Excellence Program (Project No. 2017-1.2.1-NKP-2017-00001) and by the ELTE Excellence Program (783-3/2018/FEKUTSRAT). A.K. and G.B. acknowledge funding from DFG via FLAG-ERA project ‘iSpinText’. Growth of hexagonal boron nitride crystals was supported by the Elemental Strategy Initiative conducted by the MEXT, Japan and JSPS KAKENHI Grant Numbers JP15K21722.
\end{acknowledgments}

\bibliography{ref}

\end{document}